\documentclass[pra,twocolumn,superscriptaddress]{revtex4}
\usepackage{amsmath}
\usepackage[final]{graphicx}
\newcommand{\nc}{\newcommand}
\nc{\bc}{\begin{center}}
\nc{\ec}{\end{center}}

\begin{document}
\title{Memory Effects and Scaling Properties of Traffic Flows}
\author{Bo-Sture K.\ Skagerstam}
\email{Bo-Sture.Skagerstam@whys.ntnu.no}
\affiliation{Department of Physics, The Norwegian University of
  Science and Technology, N-7491 Trondheim, Norway }
\author{Alex Hansen}
\email{Alex.Hansen@phys.ntnu.no}
\affiliation{Department of Physics, The Norwegian University of
 Science and Technology, N-7491 Trondheim, Norway }
\begin{abstract}
  \noindent Traffic flows are studied in terms of their noise of sound, which is an easily accessible experimental quantity. The sound noise data is studied making use of scaling properties of wavelet transforms and Hurst exponents are extracted. The scaling behavior is used to characterize the traffic flows in terms of scaling properties of the memory function in Mori-Lee stochastic differential equations. The results obtained provides for a new theoretical as well as experimental framework to characterize the large-time behavior of traffic flows. The present paper outlines the procedure by making use of one-lane computer simulations as well as sound-data measurements from a real two-lane traffic flow. We find the presence of conventional diffusion as well as 
$1/f$ noise in real traffic flows at large time scales.

\\

\begin{center}
EUROPHYSICS LETTERS Vol. 72 $\bullet$ Number 4 $\bullet$ pp. 513-519
\end{center}
\end{abstract}
\maketitle

\bc{
\section{INTRODUCTION}
\label{sec:introd}
}\ec

Various aspects on traffic flows has been studied in great detail over
the years (see e.g. Refs.\cite{Wolf96,Schad2000} and references cited therein). Due to the properties of vehicles and the
variability of drivers behavior, it appears quite natural that
stochastic methods is the proper theoretical framework in modelling
of traffic flows. Numerical simulations based on the
Nagel-Schreckenberg  scenario \cite{Nag1992,Schreck1995} have provided us with various insights in
the large scale behavior of traffic flows and various actual
analytical modelling has been 
discussed in great detail the literature (for some recent accounts see e.g. Refs.\cite{recent2004}). 

It has become an increasing demand in populated areas to monitor and control the
sound noise level of traffic flows. 
It is an easy experimental task to collect sound data of a traffic flow, but it is our 
understanding that such data has not until now been used to
study the properties of the traffic 
flow by itself. In the present paper we introduce what we believe is a
new and novel experimental method to study some characteristic features of
traffic flows which can be combined with the theory of stochastic 
differential equations of the Mori-Lee type \cite{mori65,lee82}  with memory effects
included.  In the course of our discussion below we will see that this
actually can be a fruitful method. By making use of  a now well established scaling analysis in terms of wavelets, one can infer self-affine properties of traffic
flows. Generic properties of Mori-Lee \cite{mori65,lee82} 
 stochastic differential
equations are then in fact compatible with the scaling properties obtained
from the data analysis. We use artificial traffic data as obtained from
one-lane computer simulations with car stopping included as well as data
from a real field measurement of an opposite two-lane traffic flow
with a common crossroad in order to test our procedure. The real traffic data exhibits short-time scaling
corresponding to  normal Einstein diffusion, a finite time $1/f$
noise (see e.g. Ref.\cite {Davidsen2002} and references therein) behavior as well as large-time  fluctuations of the sound-level corresponding to a flat power spectrum.

The paper is organized as follows. In Section
\ref{sec:mori} we recall some basic properties of stochastic differential equations of
the Mori-Lee form in terms of conventional Laplace transform
techniques. The large-time properties of fluctuations in the noise
level are expressed in terms of the short distance properties of the
Laplace transform of the memory function. 
In Section \ref{sec:data} we analyze time-series of sound measurements of traffic flows. The scaling properties of
the traffic data is based on a now established method of wavelet
scaling properties of statistical data and the extraction of the
so called Hurst scaling exponent \cite{Arnedo1995,Jones1996,Nes1998}. In Section \ref{sec:final} we summarize our results and suggest further extensions of our analysis of traffic flows.

\vspace{0.5cm}
\bc{
\section{MEMORY EFFECTS AND THE MORI-LEE EQUATION}
\label{sec:mori}
}\ec

We consider the dynamics of the traffic system in terms of the sound noise they produce. This noise will be modelled by  a second-order stochastic
differential equation for the time-dependent sound intensity $I(t)$
treated as a stochastic variable  with a space-homogeneous memory
function of the form \cite{Morgado02}
\begin{align}
\label{eq:morilee}
 \frac{dv(t)}{dt} + \int_0^t dt'\Gamma (t-t')v(t') = F(t)~~.
\end{align}
Here $v(t)=dI(t)/dt/I(t)$, and  $F(t)$ is a stochastic background. 
Eq.(\ref{eq:morilee}) can be derived from the wave-equation for 
space-homogeneous sound propagation assuming a stochastic source
of the form $F(t)I(t)$ and including a general causal form of Mori-Lee
dissipation as well as linearizing in
time-derivatives. The time integral of $v(t)$ will then be the
sound-level $L(t)=\log(I(t)/I_{0})$ with a suitable reference level
$I_{0}$ expressed in decibel, which  is an easily measurable
quantity for traffic flows. Without lack of generality the memory function $\Gamma(t)$ and
the stochastic background $F(t)$ can be assumed to be symmetric under time-reversal since
$v(-t)=-v(t)$. The fluctuations of $F(t)$ are supposed to be stationary with
a correlation compatible with the generalized fluctuation-dissipation
theorem \cite{Kubo66}, i.e.
\begin{align}
\label{labelFF}
 \langle F(t+\tau)F(\tau)\rangle = {\cal N} \Gamma (t) = {\cal N} \int_0^{\infty}d\omega \rho_F(\omega)\cos(\omega t)~~,
\end{align}
where $\rho_F(\omega)$ is the power spectrum of the noise \cite{Morgado02}.
 In terms of conventional Laplace
transform techniques one now finds that
\begin{align}
\label{eq:vv}
\langle v(t)v(t')\rangle = (\langle v^2 \rangle_0 - {\cal N})f(t)f(t') + \nonumber \\
{\cal N}(f(t-t')\Theta (t-t')+f(t'-t)\Theta (t'-t))~~.
\end{align}
Here
\begin{align}
f(t)= \frac{1}{2\pi
  i}\int_{\gamma - i\infty}^{\gamma + i\infty} ds\frac{e^{st}}{s+\Gamma (s)}
\end{align}
where $\gamma $ is a sufficiently large positive number and where
$\langle v^2 \rangle_0 \equiv \langle v(0)v(0)\rangle $. The initial data leads in general to the
normalization condition $f(0)=1$  and stationarity
of $\langle v(t)v(t')\rangle$ implies  ${\cal N} = \langle v^2 \rangle_0$ as well as $f(0)=1$.  Eq.(\ref{eq:morilee}), furthermore, leads to the following general equation for the
fluctuations of the sound-level $L(t)$:
\begin{align}
\label{eq:noise}
\Delta L(t)^2 \equiv \langle (L(t)-L(0))^2 \rangle & = \nonumber \\(\langle v^2 \rangle_0 -{\cal N})f_1(t)^2  + 2{\cal N}f_2(t)& ~~,
\end{align}
where we have defined the functions
\begin{align}
\label{eq:f1}
f_1(t)= \frac{1}{2\pi
  i}\int_{\gamma - i\infty}^{\gamma + i\infty} \frac{ds}{s}\frac{e^{st}-1}{s+\Gamma (s)}~~,
\end{align}
and
\begin{align}
\label{eq:f2}
f_2(t)= \frac{1}{2\pi
  i}\int_{\gamma - i\infty}^{\gamma + i\infty}
  \frac{ds}{s^2}\frac{e^{st}- 1- st}{s+\Gamma (s)}~~.
\end{align}
For a stationary process the fluctuations in the sound-level $L(t)$ is
determined by the second term in Eq.(\ref{eq:noise}) with a unique
short-time behavior, i.e.
\begin{align}
\Delta L(t) = t\sqrt{{\cal N}} + {\cal O}(t^2)~~,
\end{align}
and in principle the parameter ${\cal N}$ can therefore be determined uniquely by a study of the sound fluctuations at sufficiently small time-scales.
It is clear from Eqs.(\ref{eq:noise}) and (\ref{eq:f2}) that the large-time behavior of $\Delta L(t)$ is controlled by the properties of $\Gamma (s)$ at small $s$  \cite{Morgado02}. If $\Gamma(s)\simeq s^{\alpha -1}$ at small $s$, i.e. $\Gamma(t)\simeq t^{-\alpha}$ at large $t$, one predicts that $\Delta L(t) \simeq t^{\alpha/2} $, if $\alpha \leq 2$.  If e.g. $\Gamma(s)= \xi$ for all $s$,  with $ \xi $ real and positive, we obtain the conventional large-time behavior
\begin{align}
\label{eq:flat}
\Delta L(t)  = \sqrt{2Dt}~~,
\end{align}
with a diffusion constant $D={\cal N}/\xi$. If $\Gamma(s)\simeq \xi^2/s$, with $ \xi $ real and positive, corresponding to a $1/\omega $ power spectrum of $\rho_F(\omega)$ at small $\omega$, we obtain
\begin{align}
\label{eq:1/f}
\Delta L(t)  = \frac{\sqrt{2{\cal N}}}{\xi }~~,
\end{align}
apart from possible oscillatory and/or logarithmic corrections. If $\alpha < 2$ we obtain from Eq.(\ref{eq:vv}) the asymptotic limit $\langle v^2 \rangle_{\infty} \equiv \lim_{t \rightarrow \infty} \langle v(t)^2 \rangle ={\cal N}$. In the limiting case $\alpha =2$ we write $\Gamma (s) = cs$ for small $s$, where $c \neq -1$ is a constant. We then find that $ \langle v^2 \rangle_{\infty}  = (\langle v^2\rangle_0  -{\cal N})/(1+c)^2 + {\cal N}$ and $\Delta L(t)^2 = (\langle v^2 \rangle_{0} + c{\cal N})t^2/(1+c)^2$. For a stationary correlation $\langle v(t)v(t') \rangle$ we then see that ${\cal N} =\langle v^2 \rangle_{\infty} ( = \langle v^2 \rangle_{0})$ and $\Delta L(t)^2 = {\cal N}t^2/(1+c)$.  In passing we notice that for $\alpha > 2$,  Eq.(\ref{eq:vv}) leads to the asymptotic value $ \langle v^2 \rangle_{\infty}  =   \langle v^2 \rangle_0$ and according to Eq.(\ref{eq:noise}) the asymptotic fluctuations are also independent of ${\cal N}$ in this case, i.e.  $\Delta L(t)^2 = \langle v^2 \rangle_0 t^2$ for large values of $t$.  In Ref.\cite{Costa2003} it was also argued that in general ${\cal N} \neq \langle v^2 \rangle_{\infty}$, if $\alpha \geq 2$, which was interpreted as an actual breakdown of the fluctuation-dissipation theorem. The results presented  above summarizes what we need in order to analyze the large-time behavior of traffic flows. 

\bc{
\section{WAVELET ANALYSIS OF TRAFFIC FLOW DATA}
\label{sec:data}
}\ec

The new experimental method we suggest is to make use of the noise of sound as produced by traffic flows.
The large-time behavior of the sound data obtained from such traffic flows can be characterized in terms of the scaling properties of suitably averaged wavelet amplitudes of an input  signal $S(t)$, i.e.
\begin{align}
W\left[S\right](\lambda a)= \lambda ^H W\left[S \right](a)~~,
\end{align}
where $H$ is the Hurst exponent and $W\left[S\right]$ stands for the wavelet transform of the signal $S(t)$ given in form of a time-series. Our normalization is such that $H=0.5$ corresponds to 
conventional Brownian motion. The averaging procedure we make use of follows the suggestion of Ref.\cite{Nes1998}, i.e. at a given scale $a$ we average over the absolute values of  the signal $S(t)$ in the wavelet domain and hence $W[S](a) \simeq \Delta S(t)$ with $t \simeq a $. $\Delta L(t)^2$ then scales like $t^{2H}$ at large times. If we, in a similar manner, define a Hurst exponent $H_F$
of the stochastic background $F(t)$, i.e. $\langle F(t+\tau)F(\tau)\rangle$ scales like $t^{2H_F}$ at large times, then $H=-H_F =\alpha/2$ according to Eqs.(\ref{labelFF}) and (\ref{eq:noise}).
In our numerical work we have made use of the Daubechies wavelet family of order four \cite{daub92,Newland93}. 

\begin{figure}[htp]
\unitlength=0.5mm
\begin{picture}(160,140)(0,0)
\includegraphics{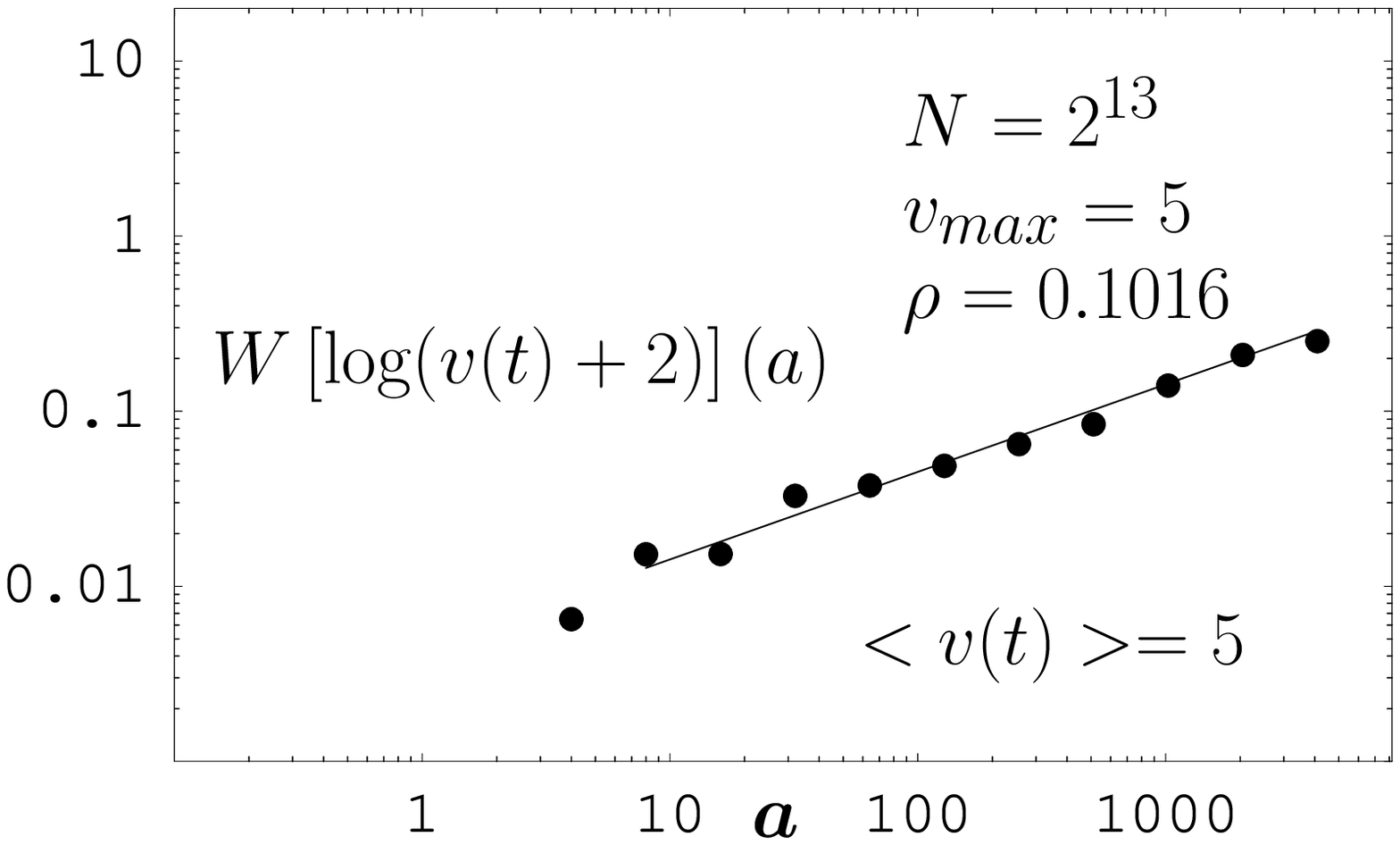}
\end{picture}
\vspace{-5mm}
\caption{\label{fig:fig11} This figure shows the typical scaling of averaged wavelet
  coefficients for the $\log ( v(t)+2) $ data as obtained from a one-lane simulation with
  a traffic intensity of $0.53$ cars per unit time at the point of
  sound measurements with
  $N=2^{13}$ data points. All identical cars moves in this case with the maximum
  speed of $5$ lattice units per unit time. The density of cars per lattice site is
  $\rho =0.1016$. For large time-scales we
  obtain an expected Hurst coefficient $H=0.5$ corresponding to normal
  Einstein diffusion.}
\end{figure}

As an example we have used data from a one-lane simulation with identical cars located at sites, with an initial occupancy probability $p$, of a lattice with $10^4$ lattice points with periodic boundary conditions and $N=2^{13}$ data points in time  in order to illustrate our method. The actual value of cars per site, i.e. the car density per lattice site $\rho$, change very little during the simulation and approaches $p$ in the limit of a large lattice. The cars have a hard core repulsion so that only one car at a time occupies a given lattice site and a speed limit restricts the car velocities. In order to simulate the approach to an optimal traffic flow, a given car either  speeds up or slows down depending on the distance to the car ahead. The velocity increases with one lattice unit per unit time if this is allowed by the distance to the car ahead. Otherwise the velocity is decreased with one lattice unit per unit time (for details of the implementation of such a simulation see e.g. Ref.\cite{Gaylord1995}). In order to make a concrete model we assume that the sound produced from a car at the site of the detection device  leads to a sound-level  $L(t)=\log (2+v(t))$, where $v(t)$ is the velocity of the car at the site in lattice units  per unit time. If there is no car present we put $v(t)=-1$, which simply corresponds to no sound at all. 
\begin{figure}[htp]
\unitlength=0.5mm
\begin{picture}(160,140)(0,0)
\includegraphics{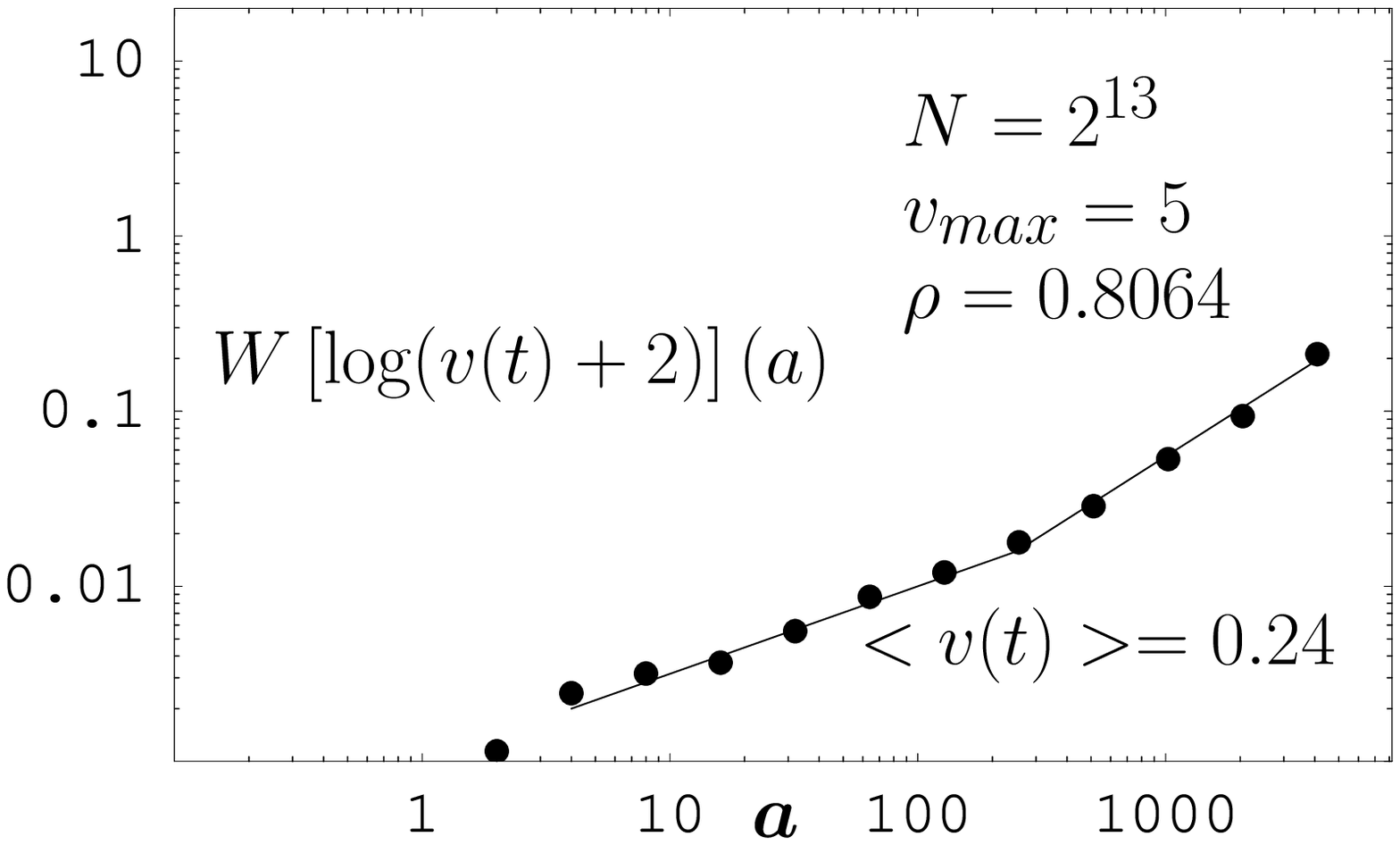}
\end{picture}
\caption{\label{fig:fig22} This figure shows the scaling of wavelet
  coefficients as in Fig.\ref{fig:fig11} but with
  a high traffic density $\rho=0.8064$ per lattice site. The average velocity is in this case $<v(t)>=0.24$ lattice units per unit time. The traffic intensity is $0.1936$ cars per unit time.   The wavelet analysis reveals two types of characteristic behavior. There is a range of a scale with an expected $H=0.5$ corresponding to normal
  Einstein diffusion and a large-time scaling with $H=0.9$ corresponding to a superdiffusive behavior. }
\end{figure}

In Fig.\ref{fig:fig11} we show the typical scaling of averaged wavelet
coefficients  for $S(t)= \log (v(t)+2)$ at a traffic intensity of $0.53$ cars per unit time. The  maximum speed of the cars is $5$ lattice units per unit time and $p=0.1$. The actual number of cars per lattice site, $\rho$, is found to be $0.1016$. For large time-scales we
obtain an expected Hurst exponent $H=\alpha/2=0.5$ corresponding to normal Einstein diffusion.  It is interesting to observe that this scaling can be modelled by a constant power spectrum $\rho_F(\omega)$, if $0 \leq \omega \leq \omega_{max}$,  with a cutoff parameter $\omega_{max}$ that still opens up a window for various finite-time behavior \cite{Morgado02}. We intend to discuss this feature of our model in more detail elsewhere. Even though the first data points corresponds to a linear $t$ scaling we do not attribute much statistical significance to this behavior. In the next simulation we increase the density of cars, i.e. we choose $p=0.8$, but keep the rest of the parameters the same. Due to the presence of traffic jamming we expect an increase of the fluctuations of noise at larger time scales and therefore a large diffusion constant, i.e. a superdiffusive behavior. Due to the larger car density one may expect a lower mean velocity and we actually  find that $<v>=0.24$ lattice units per unit time corresponding to a lower traffic intensity of $0.194$ cars per unit time as compared to the simulation of Fig.\ref{fig:fig11}.  As presented in Fig.\ref{fig:fig22} we find, as expected, a normal diffusive behavior at small scales. At  larger scales we find a Hurst exponent close to $H=1.0$, i.e. a clear superdiffusive behavior. A model for such a  superdiffusive behavior has been discussed in Ref.\cite{Morgado02} in terms of a constant power spectrum $\rho_F(\omega)$ in a finite frequency interval $0<\omega_{min }\leq \omega \leq \omega_{max}$ with parameters which again opens up a window for various finite-time behavior.

\begin{figure}[htp]
\unitlength=0.5mm
\begin{picture}(160,140)(0,0)
\includegraphics{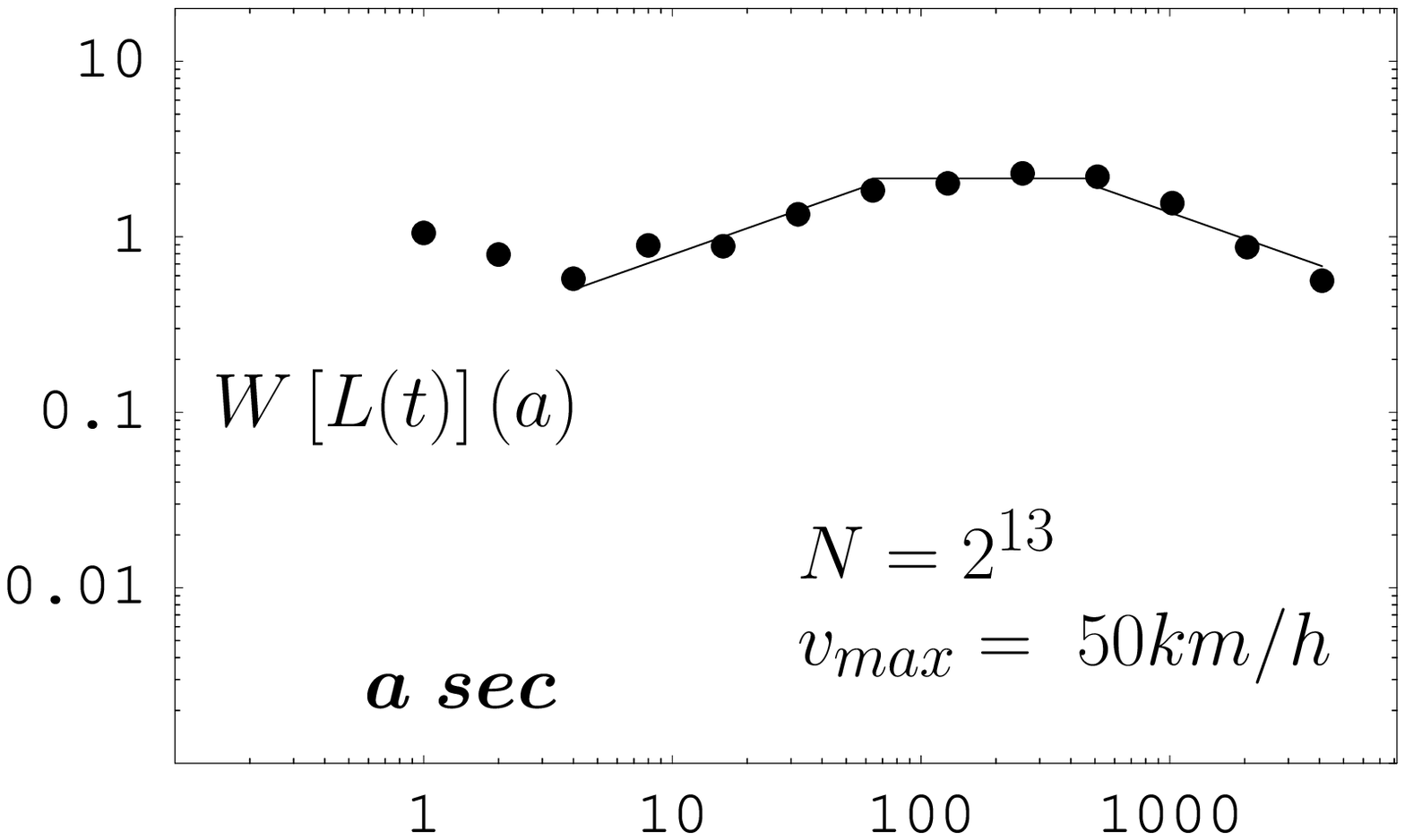}
\end{picture}
\caption{\label{fig:fig33} This figure shows the scaling of wavelet
  coefficients for the sound-level $L(t)$ using sound data from a real two-lane traffic flow with $N=2^{13}$ measurements each with a time-window of one second. The conventional diffusive behavior $H=0.5$ is clearly seen at small scales. At large scales  the sound-level becomes peaked around its mean value with a white-noise power spectrum for $\Delta F(t)^2$ which, formally, corresponds to $H=-0.5$. At intermediate scales we uncover a $1/f$-noise which indicates a self-regulatory behavior of the traffic flow.}
\end{figure}
\vspace{0.5cm}

We have also performed a field measurement of a real traffic flow during a period of a high traffic intensity.
The traffic data is based on a sample of $N=2^{13}$ measurements, one per second, on a two-lane finite road segment with opposite traffic but with a common crossroad. The required instrument is a low-cost standard decibel meter with sound data automatically stored \cite{decibel}. We expect  conventional diffusion at  short time-scales. At sufficiently large scales we expect a constant sound-level with a flat power spectrum of the sound fluctuations since the presence of the common traffic obstacle should average out in the data at large time scales. A flat power spectrum for the fluctuations of the sound-level leads to $\Delta L(t)^2 \simeq 1/t$ at large times. For medium scales it is not entirely clear in what manner the driver of the vehicles adapts to each other. As seen in Fig.\ref{fig:fig33} we find the expected small scale $H=0.5$ behavior. At large scales we find a Hurst exponent $H=-0.5$, i.e. formally $\Gamma(t) \simeq t $ at large times.  This behavior corresponds to the expected flat, i.e. white-noise, power spectrum of the fluctuations $\Delta L(t)^2$. It, however, appears that our model above breaks down for $H_F=-H > 0$, i.e. $\alpha < 0$, and we do not consider this behavior in more detail.  At intermediate scales we find a low-frequency power spectrum $\rho_F(\omega)\simeq 1/\omega$, i.e.  the famous $1/f$ noise (see e.g. Ref.\cite {Davidsen2002} and references therein), and the sound noise fluctuations are constant as in Eq.(\ref{eq:1/f}). In order to be more specific, we consider as a model for such a behavior  a power spectrum of the form $\rho_F(\omega)=B\exp(-\omega/\Omega)/\omega$, valid for $\omega\geq \omega_{min} > 0$,  where $\omega_{min}$ plays the role of an infrared cutoff and $B$ is a normalization constant. $\Omega >> \omega_{min}$ is a corresponding ultraviolet cutoff. The memory function  is then of the form $\Gamma(t) \simeq B\log(\Omega\exp(-C)/\omega_{min}(1+(\Omega t)^2)^{1/2})$, valid for times $t<\exp(-C)/\omega_{min}$. Here $C\approx 0.57721566...$ is the Euler-Mascheroni constant. One then finds that $\Gamma (s) \simeq B\omega_{min}/s\Omega$. 
The importance of a $1/f$-noise in traffic flows has been noticed before (see e.g. Refs.\cite{Chen2001,nagel95}). Here we presented a method in which such a behavior is easily revealed in sound fluctuations of traffic flows.

\vspace{0.5cm}
\bc{
\section{FINAL REMARKS}
\label{sec:final}
}\ec
\vspace{-0.5cm}

In summary, we have considered some aspects of the large-time structure of traffic flows. We have suggested a cost-effective method to collect traffic data in terms of the noise of sound the traffics flows actually produce. The scaling behavior observed can easily be modelled in terms of a general form of stochastic differential equation with memory effects included which opens up an avenue of detailed analysis.  In our presentation of the method
retardation effects, Doppler shifts,  frequency dependence of the sound and other features, which can be inferred from sound observations,  have all been neglected but can be used in more realistic situations to extract details of the monitored vehicles, flow rates and other characteristics of traffic flows.
Our data from a realistic traffic situation also suggest the presence of a $1/f$ noise, a feature which has been suggested elsewhere to be of great importance in the self-regulatory behavior of intense traffic flows \cite{nagel95}.

Even though we have applied our methods in order to model properties of traffic flows, it appears that the considerations outlined in the present paper have a much broader range of applicability. We intend to come back to such issues elsewhere.

\vspace{0.2cm}
\begin{center}
{ \bf ACKNOWLEDGMENT }
\end{center}
%

One of the authors (B.-S.S.)  wishes to thank NorFA for financial support and
 G\"{o}ran Wendin  and the Department of Microelectronics 
and Nanoscience  at Chalmers University of Technology
 and G\"{o}teborg University for hospitality. We are also indebted to the director C. Skagerstam of Procaan AB, Sweden, for assistance in the sound measurements.


\end{document}